\documentclass[11pt]{article}  
\usepackage{verbatim}
\usepackage{amssymb}      
\usepackage{amsfonts}   
\newskip\humongous \humongous=0pt plus 1000pt minus 1000pt

\newif\ifdtup

\catcode`\@=11
 
\@addtoreset{equation}{section}
 
\def\@normalsize{\@setsize\normalsize{15pt}\xiipt\@xiipt
\abovedisplayskip 14pt plus3pt minus3pt%
\belowdisplayskip \abovedisplayskip
\abovedisplayshortskip \z@ plus3pt%
\belowdisplayshortskip 7pt plus3.5pt minus0pt}
 
\def\small{\@setsize\small{13.6pt}\xipt\@xipt
\abovedisplayskip 13pt plus3pt minus3pt%
\belowdisplayskip \abovedisplayskip
\abovedisplayshortskip \z@ plus3pt%
\belowdisplayshortskip 7pt plus3.5pt minus0pt
\def\@listi{\parsep 4.5pt plus 2pt minus 1pt
     \itemsep \parsep
     \topsep 9pt plus 3pt minus 3pt}}
 
\relax

\catcode`@=12
 
\catcode`\@=11
 
\def\section{\@startsection{section}{1}{\z@}{3.5ex plus 1ex minus
   .2ex}{2.3ex plus .2ex}{\large\bf}}
 
\def\thesection{\arabic{section}.}

\def\appendix{\setcounter{section}{0}
 \def\thesection{Appendix \Alph{section}:}
 \def\theequation{\Alph{section}.\arabic{equation}}}

\begin{document}

\newcommand{\beq}{\begin{equation}}
\newcommand{\eeq}{\end{equation}}
\newcommand{\bea}{\begin{eqnarray}}
\newcommand{\eea}{\end{eqnarray}}
\newcommand{\beas}{\begin{eqnarray*}}
\newcommand{\eeas}{\end{eqnarray*}}
\newcommand{\defi}{\stackrel{\rm def}{=}}
\newcommand{\non}{\nonumber}
\newcommand{\bquo}{\begin{quote}}
\newcommand{\enqu}{\end{quote}}
\def\de{\partial}
\def\Tr{ \hbox{\rm Tr}}
\def\const{\hbox {\rm const.}}
\def\o{\over}
\def\im{\hbox{\rm Im}}
\def\re{\hbox{\rm Re}}
\def\bra{\langle}\def\ket{\rangle}
\def\Arg{\hbox {\rm Arg}}
\def\Re{\hbox {\rm Re}}
\def\Im{\hbox {\rm Im}}
\def\diag{\hbox{\rm diag}}
\def\longvert{{\rule[-2mm]{0.1mm}{7mm}}\,}
\bigskip

\title { {\footnotesize   \hfill IFUP-TH/2003/20, SISSA 52/2003/FM} \\
Calculating   Gluino Condensates in  $\mathcal{N}=1$ SYM   \\
                                from  Seiberg-Witten Curves }  
\author{Kenichi KONISHI \footnote{ konishi@df.unipi.it}\\
                    \small{\it Dipartimento di Fisica ``E. Fermi",
Universit\`a di Pisa,} \\    
\small{\it INFN, Sezione di Pisa}    \\ 
                    \small{\it Via Buonarroti, 2, Ed. C, 56127 Pisa,
Italy}                     \\
                and  \\
 Antonio RICCO \footnote{ricco@sissa.it}\\
                    \small{\it S.I.S.S.A. - I.S.A.S. International
School for Advanced Studies,} \\
                  \small{\it Via Beirut 2-4, 34014 Trieste, Italy}
             }
\date{}

\maketitle

\begin{abstract}
We determine  the gluino condensate $\left< \Tr \lambda^2 \right>$ in
the pure ${\cal N}=1$ super Yang-Mills theory  (SYM)    for the classical gauge
groups $SU(r+1)$, $SO(2r+1)$, $USp(2r)$ and  $SO(2r)$, 
by deforming the pure ${\cal N}=2$ SYM   theory
with the adjoint scalar multiplet mass, following the work 
by Finnell and Pouliot,  and  Ritz  and Vainshtein.
The value of the gluino
condensate agrees in all cases with what was found in the
weak coupling istanton calculation.
\end{abstract}

The value of the gluino condensate in
the pure $\mathcal{N}=1$ super Yang-Mills theory with gauge group
$SU(r+1)$   has been calculated   in the eighties by using
 two different methods. In the first one
\cite{Novikov:1983ek},   called  the strong coupling
instanton calculation,  one
evaluates the (r+1) point function $\left< \Tr \lambda^2 (x_1) \dots
\Tr \lambda^2 (x_{r+1})\right>$  which is saturated by one
instanton zero modes at short distances, and being constant due to
supersymmetric Ward-Takahashi identities,  it is set  equal to the 
product of  $\left< \Tr \lambda^2
\right>$ by  clustering.

The second method
\cite{Affleck:1984mk},
the so-called  weak coupling istanton calculation, makes use of a 
deformation of the original theory by addition of some matter fields. 
 Classically, the theory acquires  a flat direction, allowing one 
to compute the instanton effects reliably at large scalar vacuum 
expectation value (VEV), where the theory is weakly coupled.  {\it  
After} the calculation   one adds the 
matter mass, which eliminates the flat direction,  and the scalar VEV 
gets  fixed now  by the matter mass and $\Lambda$.   Upon  decoupling  
the matter,  by matching  the scales of the  theories with and without 
the  matter fields appropriately,  one finds   $\left< \Tr \lambda^2 
\right>$ for SYM.

These two methods yields  different results (``${4 \o 5}$ puzzle"), leading to
various
discussions \cite{Amati:1988ft, kovner}.
More recently,  a third  elegant   method \cite{Davies:1999uw} has been utilized  
in which one calculates the gaugino condensate directly in the 
semiclassical
approximation, in the spacetime compactified  on $\mathbb{R}^3 \times S^1$.
 This method gives  results in agreement with the weak coupling 
instanton calculation \cite{Affleck:1984mk, Finnell:1995dr}  
for    the classical gauge 
groups $SU(r+1)$, $SO(2r+1)$, $USp(2r)$ and  $SO(2r)$.

Still another method makes
use of  the exact solution of the $\mathcal{N}=2$   SYM  theories
\cite{Seiberg:1994rs}. One first computes the
gluino condensate in the $\mathcal{N}=2$ SYM 
perturbed by the adjoint mass term,   $\mu \, \Tr \, \Phi^2$,  and then
decouples the adjoint 
matter fields, going back to
pure   $\mathcal{N}=1$  SYM.   It has been applied to the $SU(2)$
case in 
\cite{Finnell:1995dr}.   In the spirit it is similar to the second 
method  but, instead of a semiclassical  
approximation,  it  exploits   the exact quantum sum encoded    in the 
Seiberg-Witten curves \cite{Seiberg:1994rs, Brandhuber:1995zp}.

In this paper,    this method  will be applied to  the softly broken
$\mathcal{N}=2\, $  SYM   with classical 
gauge groups
$SU(r+1)$, $SO(2r+1)$, $Sp(2r)$ and  $SO(2r)$.
The result we find   agrees in all cases   with that found in  
\cite{Affleck:1984mk,Davies:1999uw,Finnell:1995dr}.
  The agreement is    remarkable as the methods used in 
\cite{Affleck:1984mk,Davies:1999uw}  and in this
paper     involve many   different steps, and confirms   
the correctness of the weak coupling   
instanton results.   As the relation such as $\,S^N \equiv (\Tr \lambda
\lambda)^N   =  \Lambda^{3N} $   (for $SU(N)$)
  has recently been understood  as an exact operator relation in the chiral
rings \cite{CDSW},   our  result  might  not
be really surprising, but  it is reassuring,  and nontrivial,  that things work
properly.

One common feature among the method used here and those  of 
\cite{Affleck:1984mk,Davies:1999uw}  is that  
the determination of the condensates is  done in a well-defined  vacuum, 
in contrast to
the  case of the  standard ``strong coupling instanton method".   
The main  difficulty  in that approach    is the presence of a  discrete 
degeneracy of  vacua, which led the authors of \cite{Amati:1988ft} 
to  interpret   the standard instanton contribution  as providing a sum  
over these vacua, and  to apply an algorithm for  ``disentangling 
the vacuum sum", in order to get  $\left< \Tr \lambda^2 \right>$   in 
each  vacuum.  
 In view of the results found in \cite{Affleck:1984mk,
Davies:1999uw,Finnell:1995dr} 
and here such a procedure requires a careful
re-examination.  The problem is an important one  as it has to do with 
the issue of  dominant vacuum gauge-field configurations
in  a confining  theory. 
 From this point of view, it is interesting   to observe   that   in the 
approach followed here  the condensate $\bra \Tr \,\phi^2 \ket$
through which the gaugino condensate is computed,  gets  equal 
contributions from all instanton numbers, as can be
explicitly   verified   (for $SU(2)$)  by using Matone's  instanton recursion 
relations \cite{Matone}.

The auxiliary elliptic curves for the classical gauge groups are 
\cite{Seiberg:1994rs, Brandhuber:1995zp}
\begin{eqnarray}\label{eq:curves}
SU(r+1): \quad &
            y^2 = \frac{1}{4} \prod_{a=1}^{r+1} \left( x - \phi_a \right)^2
                                    - \Lambda^{2r+2}\;,\non \\
SO(2r+1):\quad &
                        y^2 = x \prod_{a=1}^{r} \left( x - \phi_a^2 \right)^2
                                    - \Lambda^{4r-2} x^2\;,\non \\
USp(2r): \quad &
                        y^2 = \prod_{a=1}^{r} \left( x - \phi_a^2 \right)
                        \left\{
                        x \prod_{a=1}^{r} \left( x - \phi_a^2 \right)
                                    - \Lambda^{2r+2}     \right\} \;,
\non  \\
SO(2r): \quad &
                        y^2 = x \prod_{a=1}^{r} \left( x - \phi_a^2
\right)^2
                                    - \Lambda^{4r-4}x^3 \;.
\end{eqnarray}
In our notation, at a classical level we have $\phi = \diag \left(  
\phi_1, \dots ,\phi_{r+1} \right)$ for $SU(r+1)$
and $\phi = \sum_{a=1}^{r}\phi_a H^a$ for the other groups, where $H^a$
are the generators of the Cartan subalgebra, with the normalization
\begin{equation}
\Tr H^a H^b = \delta^{a b} \;.
\end{equation}
{ We proceed as follows: we first
  {(i)} determine the vacua of the deformed theory;
  {(ii)} find   the relation between the parameter $\Lambda$ appearing 
in the
Seiberg-Witten  curve and
the scale of the theory $\Lambda_{\mathcal{N}=2}$;
 {(iii)} evaluate  $\left< \Tr \phi^2 \right>$ in each  ${\cal N}=1$ 
vacuum by using the results of (i) and (ii);
{(iv)} calculate  $\left< \Tr \lambda^2 \right>$ by using the Konishi
anomaly \cite{Konishi:1984};  then finally
{(v)}  determine  the gluino condensate in the pure ${\cal N}=1 $  limit 
($\mu \rightarrow\infty$) by matching
 the
dynamical scales $\Lambda_{\mathcal{N}=2} $ to
$\Lambda_{\mathcal{N}=1}$.}

As for the first point
the vacua of the deformed theory are those where the monopoles of  the 
maximal abelian effective gauge group $U(1)^r$ 
are all   massless, i.e. the points where the curve has the maximal 
number of
double zeros.  One must
satisfy  $r$ indipendent equations imposing pairwise equality between
the zeros.
These polynomial equations are generally difficult  to solve.
This problem can however be  sidestepped   as in 
\cite{Douglas:1995nw,Brandhuber:1995zp, Carlino:2000uk}
by making use of  the properties of the Chebyshev polynomials.

In the case of the gauge group $SU(r+1)$ \cite{Douglas:1995nw}    the factor
 $\prod_a (x-\phi_a)$ in the curve (\ref{eq:curves}) is identified  with 
a Chebyshev
polynomial of the first kind $T_{r+1}$. There are
$2(r+1)$ possible such
choices, but,  because of the Weyl invariance, only one half of them
generates different vacua:
\begin{eqnarray}
            y^2 \equiv
                    \Lambda^{2r+2}     \left\{ T_{r+1}^2\left(\xi\right)
-1 \right\}
            \;,         \qquad
            \xi=\frac{\displaystyle x}{\displaystyle 2 \Lambda} \, e^{
-  2 \pi i \,k  / 2  (r+1)} \,.
\end{eqnarray}
The right hand side  (of (\ref{eq:relazioneTU})) has two 
simple
zeros and $r$ double zeros.
The corresponding values of $\phi_a$ are
\begin{eqnarray}
    \phi_a = 2 \, \Lambda \, \omega_a^{[r+1]} \,e^{ 2 \pi i \,k  / 2 
(r+1)}\;, \qquad a = 1, \dots, r+1 \;,
\end{eqnarray}
where $\omega_a^{[n]}$ are the zeros of the Chebyshev polynomial  of the
first kind and degree $n$. Note that $\sum_{a} \phi_a=0$.

For $SO(2r+1)$, after imposing $\phi_r=0$ one identifies the term
$\sqrt{x}\prod_{a=1}^{r-1} (x-\phi_a^2)$ with a polynomial $T_{2r-1}$ in
$\sqrt{x}$:
\begin{eqnarray}
          y^2 \equiv
                  \Lambda^{4r+2} \left( \frac{\displaystyle
x^2}{\displaystyle \Lambda^4}\right)
                                                             \left\{
T_{2r-1}^2\left(\xi\right) -1\right\}   
   \;, \qquad
            \xi= 2^{\frac{1}{2r-1}}\frac{\displaystyle
\sqrt{x}}{\displaystyle 2 \Lambda} 
\;                                                                       
                        e^{ -  2 \pi i \,k  / 2  (2r-1)} \;.
\end{eqnarray}
The curve has $r$ double zeros and one single zero (as a polynomial in
$x$) and
\begin{eqnarray}
    \phi_a &=& 2^{1-{\frac{1}{2r-1}}} \,\Lambda \, \omega_a^{[2r-1]} e^{ 2
\pi i \,k  / 2  (2r-1)}                                                          \;, \qquad 
a = 1, \dots, r-1 \;,  \non \\
    \phi_r &=& 0 \;.
\end{eqnarray}

For $USp(2r)$   it is necessary to distinguish two cases. If $r$ is 
even,  one imposes
$\phi_{2a-1}=\phi_{2a}$ and then, as for $SO(2r+1)$, identifies
$\sqrt{x}\prod_{a=1}^{r/2} (x-\phi_{2a}^2)$
with a polynomial $T_{r+1}$ in $\sqrt{x}$  (see the second of
Eq.~(\ref{identify1})): 
\begin{eqnarray}
                   y^2 &\equiv &
                           \Lambda^{4r+2} \left( 2^{\frac{2r^2}{r+1}}
\right)
                                                        \prod_{a=1}^{r/2}
                                                        \left[  \xi^2 -
(\omega_a^{[r+1]})^2 \right]^2
                                                         \left\{
T_{r+1}^2\left(\xi\right) -1\right\} \;,     \nonumber \\
                    \xi&=&  2^{\frac{1}{r+1}} \frac{\displaystyle
\sqrt{x} }{\displaystyle  2 \Lambda} \,    e^{- 2 \pi i\, k / 2 (r+1)}
\end{eqnarray}
Also in this case the curve has $r$ double zeros and one simple zero.
The values of $\phi_a$ are
\begin{eqnarray}
    \phi_{2a-1} = \phi_{2a} = 2^{1-{\frac{1}{r+1}}} \Lambda \,
\omega_a^{[r+1]} \,    e^{2 \pi i\, k / 2 (r+1)}\;, \qquad a = 1, \dots,
\frac{r}{2}\;.
\end{eqnarray}

For $USp(2r)$  with odd $r$, one identifies $x_0=\phi_r^2$ with  the  
simple
zero and, after the shift $\widetilde{x}= x- \phi_r^2/2$, one identifies
the term
\begin{eqnarray}
    \left(\widetilde{x}^2- \frac{\phi_r^4}{4} \right)
    \prod_{a=1}^{r-1} \left( \widetilde{x}-\phi_{a}^2 +
\frac{\phi_r^2}{2} \right)^2
\end{eqnarray}
with a polynomial of the form $(x^2 - 1) U_{r-1}^2 (x)$
(Eq.~(\ref{eq:relazioneTU})). In this way the curve becomes 
\begin{eqnarray}
                   y^2 &\equiv &
                           \Lambda^{4r+2}
                                                        \left(
(-1)^{-\frac{1}{r+1}} 2^{1-\frac{2}{r+1}} \right)
                                                        \left( \xi - 1
\right)                                                       
U_{\frac{r-1}{2}}^2\left(\xi\right)                                                      
T_{\frac{r+1}{2}}^2\left(\xi\right)     \;,
                                                        \nonumber \\
                    \xi &=&  -  2^{\frac{2}{r+1}-1}                   
                                     \left(\frac{\displaystyle 
x}{\Lambda^2}-\frac{\phi_r^2}{2 \Lambda^2}
\right)
                                                         e^{ -2 \pi i\,
k /(r+1)}
\end{eqnarray}
and $\phi_a$'s  are
\begin{eqnarray}
    \phi_{2a-1}^2 &=& \phi_{2a}^2 =  2^{1-\frac{2}{r+1}} \Lambda^2
                                    \left(1+ \zeta_a^{\left[
\frac{r-1}{2} \right]}\right) e^{2 \pi i\, k /(r+1)}
                                     \;,  \qquad   a = 1, \dots,
\frac{r-1}{2}   \nonumber \;, \\
  \phi_r^2 &=&  2^{2-\frac{2}{r+1}} \, \Lambda^2 e^{2 \pi i\, k /(r+1)} \;,
\end{eqnarray}
where $\zeta_a^{[n]}$ are the zeros of the Chebyshev polynomial of the
second kind and degree $n$. The polynomial obtained in this way has one
simple zero and $r$ double zeros.

For  $SO(2r)$, the maximally singular curve can be obtained
\cite{Brandhuber:1995zp} by   imposing $\phi_r=0$ and identifying, as 
for the
other groups, the term containig the products $(x-\phi_a^2)$ with an
appropriate   Chebyshev polynomial of the first kind:
\begin{eqnarray}
                   y^2 \equiv
                           \Lambda^{4r+2} \left( \frac{\displaystyle
x^3}{\displaystyle \Lambda^6}\right)
                                                         \left\{
T_{2r-2}^2\left(\xi\right) -1 \right\},    \qquad
                    \xi= 2^{\frac{1}{2r-2}} \frac{\displaystyle
\sqrt{x}}{\displaystyle 2 \Lambda}
                                    e^{- 2 \pi i \,k  / 2  (2r-2)} \;.
\end{eqnarray}
The right hand side  contains one simple zero, one quadruple zero (at 
$x=0$) and
$r-2$ double zeros. The value of $\phi_a$ is  
\begin{eqnarray}
    \phi_a &=& 2^{1-{\frac{1}{2r-2}}} \,\Lambda \, \omega_a^{[2r-2]}     e^{2
\pi i \,k  / 2  (2r-2)}
                                     \;, \qquad a = 1, \dots, r-1 \;, 
\non \\
    \phi_r &=& 0 \;.
\end{eqnarray}

As for the second step  of our program,
the relation between $\Lambda$ and $\Lambda_{\mathcal{N}=2}$ has been
determined for the gauge group $SU(2)$ by \cite{Finnell:1995dr},
matching the perturbative coupling with the exact one, obtained from the
curve.
For the other gauge groups this relation has been found  by Ito and
Sasakura \cite{Ito:1996qj} by
considering   large expectation values of the scalar field, breaking the
gauge symmetry down to $SU(2)$ and matching the behaviour under this
breaking of both $\Lambda$ and $\Lambda_{\mathcal{N}=2}$.  It reads
\begin{eqnarray}
SU(r+1):     \quad \Lambda^2 &=& 2^{-1} \Lambda_{\mathcal{N}=2}^2 \;,\non \\
SO(2r+1):   \quad \Lambda^2 &=&
2^{\frac{7-2r}{2r-1}}\Lambda^2_{\mathcal{N}=2}\;,\non \\
USp(2r):     \quad \Lambda^2&=& \Lambda_{\mathcal{N}=2}^2\;, \non \\
SO(2r):     \quad \Lambda^2 &=&
2^{\frac{8-2r}{2r-2}}\Lambda_{\mathcal{N}=2}^2\;.
\end{eqnarray}

The third step  is  the calculation of $\left< \Tr \phi^2
\right>$. For the group  $SU(r+1)$ we obtain 
\begin{eqnarray}
    \left< \Tr \phi^2 \right>_k &=& \sum_{i=1}^{r+1} \phi_i^2 =
                                  4\, \Lambda^2 \,e^{ 2 \pi i \,k  / 
(r+1)}  \sum_{i=1}^{r+1} (\omega_a^{[r+1]})^2 = \nonumber \\
                              &=& 2 \,  (r+1) \, \Lambda^2 \,e^{ 2 \pi i \,k 
/  (r+1)}  = (r+1)\, \Lambda_{\mathcal{N}=2}^2 \,e^{ 2 \pi i \,k  / 
(r+1)}  \; \label{phisquare}
\end{eqnarray}
(cfr. Eq.~(\ref{eq:formula1})).  Note that $    \left< \Tr \phi^2
\right> $  is invariant under ${\mathbb Z}_2$  of ${\mathbb Z}_{2(r+1)}$
so that one finds $T_G=  r+1$
distinct ${\cal N}=1$ vacua. 
Analogously, we find  for the other classical groups
\begin{eqnarray}
    SO(2r+1): \left< \Tr \, \phi^2 \right>_k  &=&  (2r-1)
2^{\frac{4}{2r-1}-1}\Lambda_{\mathcal{N}=2}^2\, e^{2 \pi i\, k / T_G} 
\;, \non \\
USp(2r):   \left< \Tr \,\phi^2 \right>_k  &=& (r+1) \, 2^{1-
\frac{2}{r+1}} \, \Lambda_{\mathcal{N}=2}^2 \, e^{2 \pi i\, k / T_G} \;, 
\non  \\
    SO(2r): \left< \Tr \, \phi^2 \right>_k  &=& (2r-2)
2^{\frac{2}{r-1}-1} \Lambda_{\mathcal{N}=2}^2\, e^{2 \pi i\, k / T_G}
\;,\label{phisquare2}
\end{eqnarray}
where we used the formulas Eqs.~(\ref{eq:formula1}), (\ref{eq:formula2}) 
for
the zeros of Chebyshev polynomials and $k$ runs from $1$ to $T_G$. $T_G$
indicates the dual Coxeter number of the  gauge group considered:
$T_G= 2 r -1,\,\,   r+1, \,\, 2r -2 $,   for    $ SO(2r+1)$, $USp(2r)$ and $SO(2r)$,
respectively.    

As is well known, the introduction of adjoint mass  $\mu \, \Tr \Phi^2|_F$ 
eliminates all other vacua
leaving these   $T_G$ vacua.
 The  anomaly relation  \cite{Konishi:1984}
\begin{equation}\label{eq:UsoAnomaliaK}
\left< \frac{\Tr \lambda^2}{16 \pi^2} \right>    = \frac{\mu }{T_G}\left<
\Tr \phi^2\right> \;,
\end{equation}
is valid at all values of $\mu$;  
 by matching the dynamical scales as
$ \Lambda_{\mathcal{N}=1}^3 =  \mu  \,  \Lambda_{\mathcal{N}=2}^2$ (which is
valid in all cases  considered
here)  upon decoupling the adjoint matter, one finds 
\begin{eqnarray}
    \left< \frac{\Tr \lambda^2}{16 \pi^2} \right>_{SU(r+1)}   &=&
\Lambda_{\mathcal{N}=1}^3\;, \non \\
    \left< \frac{\Tr \lambda^2}{16 \pi^2} \right>_{SO(2r+1)}    &=&
2^{\frac{4}{2r-1}-1} \Lambda_{\mathcal{N}=1}^3\;, \non \\
    \left< \frac{\Tr \lambda^2}{16 \pi^2} \right>_{USp(2r)}      &=&
2^{1- \frac{2}{r+1}} \Lambda_{\mathcal{N}=1}^3\;, \non  \\
    \left< \frac{\Tr \lambda^2}{16 \pi^2} \right>_{SO(2r)}      &=&
2^{\frac{2}{r-1}-1} \Lambda_{\mathcal{N}=1}^3 \;,\label{result}
\end{eqnarray}
up to the phase factor $e^{2 \pi i\, k / T_G}$ that distinguishes
 the $T_G$  vacua, in agreement with
\cite{Affleck:1984mk,Davies:1999uw,Finnell:1995dr}.

\section*{Acknowledgements}

The authors acknowledge  useful discussions  with A. Ritz and A. Yung.


\appendix
\section{The Chebyshev polynomials}\label{sec:chebyshev}

The Chebyshev polynomials of the first kind are:
\begin{eqnarray}
 T_n (x)  &\equiv&  \cos \left[ n (\arccos x) \right]   \nonumber  \\
                    & =    &     \frac{n }{2} \sum_{k=0}^{[n/2]}
\frac{(-1)^{k}}{n-k} {n-k \choose k}
                                        (2x)^{n-2k}
\end{eqnarray}
where $[x]$ is the integer part of $x$; the Chebyshev polynomials of the
second kind are:
\begin{eqnarray}
U_n (x) & \equiv & \frac{\displaystyle \sin \left[ (n+1) (\arccos x)
                         \right]}{\displaystyle \sqrt{1-x^2}} \nonumber \\
                    & =    &     \sum_{k=0}^{[n/2]} (-1)^{k} {n-k \choose k}
                                        (2x)^{n-2k}
\end{eqnarray}
>From the trigonometric relation $\cos^2\vartheta+\sin^2\vartheta=1$:
\begin{equation}\label{eq:relazioneTU}
T_n^2(x) -1 = (x^2 - 1) U_{n-1}^2 (x)
\end{equation}
The zeros of $T_n$ and $U_n$ are respectively:
\begin{eqnarray}
        T_n: \qquad
                    \omega_a &=& \cos \left\{ \frac{\pi}{n}
\left(a-\frac{1}{2}\right) \right\} \nonumber \\ 
        U_n: \qquad
                    \zeta_a  &=& \cos \left\{ \frac{\pi}{n+1}a \right\}
\qquad a=1, \dots, n
\end{eqnarray}
It is useful to notice that $T_n$ and $U_n$ are of the form:
\begin{eqnarray}
    T_n(x) &= &
        \left\{
        \begin{array}{ll}
                        \frac{1}{2} \prod_{a=1}^{n/2} \left( 4 x^2 - 4
\omega_a^2 \right) &
                        \textrm{(n even)}   \\
                            \frac{1}{2} (2x) \prod_{a=1}^{(n-1)/{2}}
\left( 4 x^2 - 4 \omega_a^2 \right) &
                            \textrm{(n odd)}
        \end{array} \right.     \label{identify1}\\
    U_n(x) &= &
        \left\{
        \begin{array}{ll}
                        \prod_{a=1}^{n/2} \left( 4 x^2 - 4 \zeta_a^2
\right) &
                        \textrm{(n even)}   \\
                            2x \prod_{a=1}^{(n-1)/{2}} \left( 4 x^2 - 4
\zeta_a^2 \right) &
                            \textrm{(n odd)}
        \end{array} \right.  \label{identify2}
\end{eqnarray}
For this reason also $T_{2n}(\sqrt{x})$ and $U_{2n}(\sqrt{x})$ are
polynomials in $x$.
Finally we have the following relations for the zeros of $T_n$:
\begin{equation}\label{eq:formula1}
\sum_{a=1}^n \omega_a^2 = \sum_{a=1}^n \cos^2 \left\{ \frac{\pi}{n}
\left(a-\frac{1}{2}\right) \right\} = \frac{n}{2}
\end{equation}   
\begin{equation}\label{eq:formula2}
\sum_{a=1}^{\left[n/2\right]} \omega_a^2 = \frac{1}{2} \sum_{a=1}^n
\omega_a^2 = \frac{n}{4}
\end{equation}


\begin{thebibliography}{100}


\bibitem{Novikov:1983ek}
V.~A. Novikov, M.~A. Shifman, A.~I. Vainshtein, M.~B. Voloshin   and
  V.~I. Zakharov,
\newblock {\em Nucl. Phys.}   B229 (1983) 394;  
\, 
G.~C. Rossi and G.~Veneziano,
\newblock {\em Phys. Lett.}   B138 (1984) 195; 
\, 
D.~Amati, G.~C. Rossi  and G.~Veneziano,
\newblock {\em Nucl. Phys.}   B249 (1985) 1.


\bibitem{Affleck:1984mk}
I. Affleck, M.  Dine   and N.   Seiberg,
\newblock {\em Nucl. Phys.}   B241  (1984) 493; 
\,  
V.~A. Novikov, M.~A. Shifman, A.~I. Vainshtein, and V.~I. Zakharov,
\newblock {\em Nucl. Phys.}   B260  (1985) 157; 
\, 
M.~A. Shifman and A.~I. Vainshtein,
\newblock {\em Nucl. Phys.}, B296   (1988) 445;
\, 
A.~Yu. Morozov, M.~A. Olshanetsky, and M.~A. Shifman,
\newblock {\em Nucl. Phys.}  B304   (1988) 291.


\bibitem{Amati:1988ft}
D.~Amati, K.~Konishi, Y.~Meurice, G.~C. Rossi and G.~Veneziano,
\newblock {\em Phys. Rep.}   162   (1988) 169.    


\bibitem{kovner}
A.~Kovner and M.~A. Shifman,
\newblock {\em Phys. Rev.}  D56  (1997) 2396  {\tt [hep-th/9702174]};
\,
T.~J. Hollowood, V.~V. Khoze, W.--J. Lee and M.~P. Mattis,
\newblock {\em Nucl. Phys.} B570 (2000) 241  {\tt [hep-th/9904116]};
\,
A. Ritz and A.~I. Vainshtein,
\newblock {\em Nucl. Phys.} B566 (2000) 311  {\tt [hep-th/9909073]}.


\bibitem{Davies:1999uw}
N.~M. Davies, T.~J. Hollowood, V.~V. Khoze   and M.~P. Mattis,
\newblock {\em Nucl. Phys.} B559 (1999) 123  {\tt [hep-th/9905015]};
\,
N.~M. Davies, T.~J. Hollowood and V.~V. Khoze,
\newblock {\tt [hep-th/0006011]}. 


\bibitem{Finnell:1995dr}  
 D.~Finnell and P.~Pouliot, 
\newblock {\em Nucl. Phys.}  B453 (1995) 225  {\tt [hep-th/9503115]}.

 
\bibitem{Seiberg:1994rs}
N.~Seiberg and E.~Witten,  
\newblock {\em Nucl. Phys.}  B426  (1994) 19  {\tt [hep-th/9407087]}; 
\,
N.~Seiberg and E.~Witten,
\newblock {\em Nucl. Phys.}  B431 (1994) 484  {\tt [hep-th/9408099]};
\,
P.~C. Argyres and A.~E. Faraggi,
\newblock {\em Phys. Rev. Lett.}  74 (1995) 3931  {\tt [hep-th/9411057]};
\,
A.~Klemm, W.~Lerche, S.~Yankielowicz, and S.~Theisen,
\newblock {\em Phys. Lett.}   B344  (1995) 169  {\tt [hep-th/9411048]};
\, 
A.~Klemm, W.~Lerche, and S.~Theisen, 
\newblock {\em Int. J. Mod. Phys.} A11 (1996) 1929  {\tt [hep-th/9505150]};
\,
U.~H. Danielsson and B. Sundborg,
\newblock {\em Phys. Lett.}  B358 (1995) 273  {\tt [hep-th/9504102]};
\,
P.~C. Argyres and A.~D. Shapere, 
\newblock {\em Nucl. Phys.}  B461  (1996) 437  {\tt [hep-th/9509175]}.


\bibitem{Brandhuber:1995zp}
A.~Brandhuber and K.~Landsteiner,  
\newblock {\em Phys. Lett.}  B358 (1995) 73  {\tt [hep-th/9507008]}.


\bibitem{CDSW}  
F.~Cachazo, M.~R.~Douglas, N.~Seiberg and E.~Witten, 
\newblock {\em JHEP}   0212:071 (2002)  {\tt [hep-th/0211170]}.
  

\bibitem{Matone}   
M.~Matone, 
\newblock {\em Phys. Lett.}, B357 (1995) 342 {\tt [hep-th/9506102]}.  


\bibitem{Konishi:1984}  
K.~Konishi,  
\newblock  {\em Phys. Lett.}  B135  (1984) 439;
\,
K.~Konishi and  K.~Shizuya, 
\newblock  {\em Nuovo Cim.}  A90 (1985) 111.
  

\bibitem{Douglas:1995nw}
M.~R. Douglas and S.~H. Shenker,
\newblock {\em Nucl. Phys.} B447 (1995) 271  {\tt [hep-th/9503163]}.


\bibitem{Carlino:2000uk}
G. Carlino, K. Konishi and H. Murayama,
\newblock {\em Nucl. Phys.} B590 (2000) 37  {\tt [hep-th/0005076]};
\, 
G. Carlino, K. Konishi, S.~P. Kumar and H. Murayama,
\newblock {\em Nucl. Phys.} B608 (2001) 51  {\tt [hep-th/0104064]}. 


\bibitem{Ito:1996qj}
K. Ito and N. Sasakura,
\newblock {\em Phys. Lett.} B382 (1996) 95  {\tt [hep-th/9602073]};
\, 
K. Ito and N. Sasakura,
\newblock {\em Nucl. Phys.} B484 (1997) 141  {\tt [hep-th/9608054]}.

\end{thebibliography}
\end{document}